\documentclass[aps,prl,superscriptaddress,reprint]{revtex4-1}
\usepackage{graphicx}
\usepackage{amsmath}
\usepackage{amssymb}
\usepackage{amsfonts}
\usepackage{filecontents}
\usepackage{color}
\usepackage[normalem]{ulem} 
\usepackage{soul}
\usepackage{epstopdf}
\usepackage{fancyhdr}
\usepackage{float}

\definecolor{gold}{rgb}{0.85,.66,0}

\begin{document}

\title{Ultrafast Depopulation of a Quantum Dot by LA-phonon-assisted Stimulated Emission}

\author{F. Liu}\email[Email: ]{fengliu@sheffield.ac.uk.}
\author{ L. M. P. Martins}
\author{A. J. Brash}
\affiliation{Department of Physics and Astronomy, University of Sheffield, Sheffield, S3 7RH, United Kingdom}
\author{A. M. Barth}
\affiliation{Institut f\"{u}r Theoretische Physik III, Universit\"{a}t Bayreuth, 95440 Bayreuth, Germany}
\author{ J.~H.~Quilter}
\affiliation{Department of Physics and Astronomy, University of Sheffield, Sheffield, S3 7RH, United Kingdom}
\affiliation{Department of Physics, Royal Holloway, University of London, Egham, TW20 0EX, United Kingdom}
\author{V. M. Axt}
\affiliation{Institut f\"{u}r Theoretische Physik III, Universit\"{a}t Bayreuth, 95440 Bayreuth, Germany}
\author{M. S. Skolnick}
\affiliation{Department of Physics and Astronomy, University of Sheffield, Sheffield, S3 7RH, United Kingdom}
\author{A. M. Fox}
\affiliation{Department of Physics and Astronomy, University of Sheffield, Sheffield, S3 7RH, United Kingdom}

\begin{abstract}

We demonstrate ultrafast \emph{incoherent} depopulation of a quantum dot from above to below the transparency point using LA-phonon-assisted emission stimulated by a red-shifted laser pulse. The QD is turned from a weakly vibronic system into a strongly vibronic one by laser driving which enables the phonon-assisted relaxation between the excitonic components of two dressed states. The depopulation is achieved within a laser pulse-width-limited time of 20~ps and exhibits a broad tuning range of a few meV. Our experimental results are well reproduced by path-integral calculations.

\end{abstract}

\maketitle

The exciton-phonon coupling in semiconductor quantum dots (QDs) has attracted much interest due to its importance both in fundamental physics and in semiconductor-based quantum technologies \cite{Muller2015,Muller2015a, Majumdar2011,Majumdar2012a}. It has been known for some time that QDs can be excited by phonon-assisted transitions when the laser is detuned within the phonon sideband \cite{Weiler2012}. Very recently, it has been shown that population inversion can be achieved under these incoherent pumping conditions in the strong driving regime \cite{Petta2004, Quilter2015, Stace2005, Colless2014, Glassl2013, Hughes2013, Ardelt2014, Bounouar2015}. However, the opposite process in which an exciton in an inverted QD is de-excited with the assistance of longitudinal acoustic (LA) phonons has yet to be investigated. This process could be important for the development of tunable single QD lasers \cite{Nomura2010,Rafailov2007} or of QD single-photon sources in cavities with improved timing jitter \cite{Heinze2015,  Portalupi2015, Heinze2015, Portalupi2015, Muller2015}.

\begin{figure}
\includegraphics[width=\linewidth]{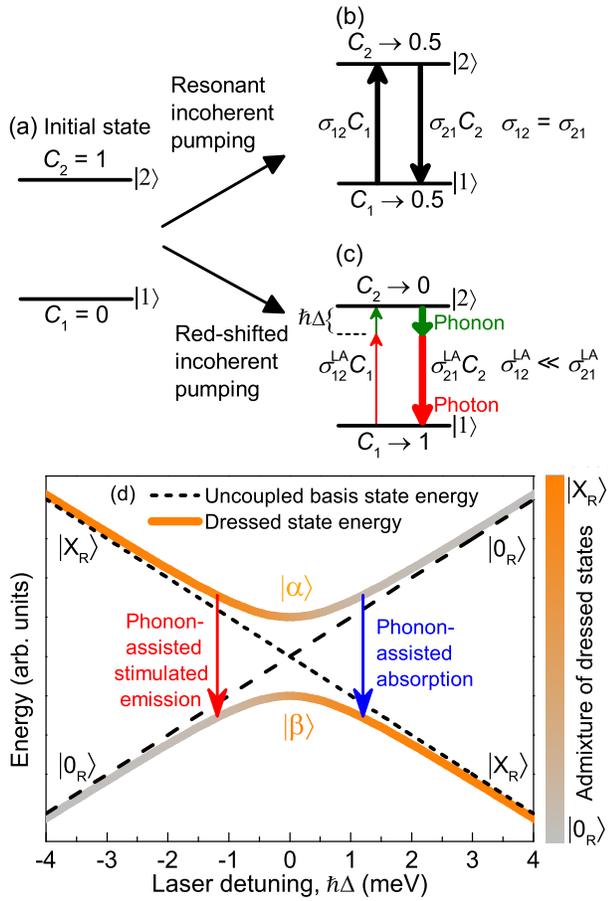}
\caption{ (a-c): Difference between resonant incoherent pumping and red-shifted incoherent pumping. (a) A 2-level system is initially inverted ($C_2>C_1$). $C_i$ denotes the population of the $i^{\rm{th}}$ state. (b) Resonant incoherent pumping induces absorption and SE with equal cross sections ($\sigma_{12}=\sigma_{21}$), moving the system towards transparency ($C_{1} = C_{2}$). $\sigma$ denotes the absorption/stimulated emission cross section. (c) If pumped by red-shifted incoherent excitation, $C_2$ can be almost completely depleted as the red-shifted laser mainly induces SE ($\sigma^\text{LA}_{12}\ll\sigma^\text{LA}_{21}$). Absorption hardly occurs due to the lack of additional phonons at low temperature. (d) The mechanism of the LA-phonon-assisted SE and absorption explained in the dressed-state picture. $|0_\text{R}\rangle$, $|X_\text{R}\rangle$: uncoupled ground state and exciton state viewed in the rotating frame. $|\alpha\rangle$, $|\beta\rangle$: optically dressed states split by the effective Rabi energy $\hbar \Lambda$.}
\label{fig1}
\end{figure}

The significance of the LA phonon-assisted de-excitation process can be appreciated by comparing it to the conventional understanding of inverted 2-level systems. Consider a quantum dot in which population inversion has been created as shown in Fig.~\ref{fig1}(a). A resonant laser pulse that is short compared to the spontaneous emission time, but long compared to the coherence time will only drive the system towards the transparency point, never crossing it due to the equal cross sections of stimulated emission and absorption [see Fig.~\ref{fig1}(b)]. However, if the dot is coupled to the lattice, stimulated emission can be induced by a red-shifted pulse via phonon emission, provided that the temperature is low enough that phonon absorption is weak [see Fig.~\ref{fig1}(c)]. The decoupling of stimulated emission and absorption enables depopulation of the inverted 2-level system to below the transparency point before spontaneous emission occurs, which is impossible for exactly resonant excitation in the incoherent limit. This process is fundamentally different to conventional vibronic systems - e.g. Ti:sapphire \cite{Klein2010} - in that the phonon coupling is weak at low light intensities, and only becomes effective at high optical powers during a strong laser pulse. In this sense, the new mechanism can be regarded as de-excitation by \textit{dynamic} vibronic coupling.

In this Rapid Communication, we demonstrate the LA-phonon-assisted stimulated emission (LAPSE) process by using a red-shifted laser pulse to \textit{incoherently} de-excite an inverted quantum dot to below the transparency point. The time dynamics indicate that the depopulation occurs in 20 ps, limited by the laser pulse width. We show that the process occurs over a broad tuning range of $\sim4$~meV (in contrast to the fixed frequency of resonant excitation) and that the results are in a good agreement with path-integral calculations \cite{Vagov2011}. The new mechanism provides a route to ultrafast reset of an excited quantum dot, which should be important for quantum photonic systems incorporating dots as the nonlinear element.

Our experiments are performed on a device consisting of InGaAs QDs embedded in the intrinsic region of an $n-i-$Schottky diode. The sample is held at $T=$ 4.2~K in a helium bath cryostat. The excitation laser pulse  is derived by spectral shaping of the output from a mode-locked Ti:sapphire laser with repetition rate 76.2 MHz. The spectral FWHM is 0.2 meV for the $\pi$ pulse in all measurements and 0.2 or 0.42 meV for the red-shifted laser pulse, corresponding to a Fourier transform limited pulse duration $\tau_\text{L}$ of 16.8 or 8~ps. $\tau_\text{L}$  is defined as the FWHM of the electric field envelope. The exciton population created by the circularly polarized laser pulse is determined by measuring the photocurrent (PC) generated when a reverse bias voltage is applied to the diode~\cite{Zrenner2002}. More details of the sample and experimental setup can be found in refs.~\onlinecite{Brash2015, Godden2012a, Kolodka2007a}.

A phenomenological description of the LAPSE process is as follows. Assume a QD is initially in the exciton state. A strong laser pulse red-shifted relative to the exciton state by $\hbar\Delta$ leads to the recombination of the exciton and the emission of a phonon and photon illustrated by thick down arrows in Fig.~\ref{fig1}(c). To fully understand the underlying mechanism, we use the dressed state picture [see Fig.~\ref{fig1}(d)]. In the rotating frame, the crystal ground state $|0\rangle$ and the incident laser field are treated as one state: $|0_\text{R}\rangle$. The exciton state $|X\rangle$ and the laser field with one photon less are treated as $|X_\text{R}\rangle$.  The relative energies of the two states are shown by the dashed/dotted lines in Fig.~\ref{fig1}(d) as a function of  the laser detuning $\hbar\Delta=\hbar(\omega_X-\omega_\text{L})$, where $\omega_X$ and $\omega_\text{L}$ are the angular frequencies of the exciton transition and laser, respectively. Since $|0_\text{R}\rangle$ and $|X_\text{R}\rangle$ are admixed by the laser field, the eigenstates of the system become two new optically dressed states: $|\alpha\rangle$ and $|\beta\rangle$, split by the effective Rabi energy $\hbar \Lambda(t)=\hbar \sqrt{\Delta^2+\Omega(t)^2}$, where $\Omega$ is the Rabi frequency for resonant excitation proportional to the electric field amplitude $E$. The dressed state energies are shown by the thick solid lines in Fig.~\ref{fig1}(d) and the color gradient illustrates the excitonic contribution to the corresponding states. Owing to their excitonic components, $|\alpha\rangle$ and $|\beta\rangle$ are coupled by LA-phonons via the deformation potential \cite{SMnote}\nocite{Breuer2002}.

\begin{figure*}
	\includegraphics[width=\textwidth]{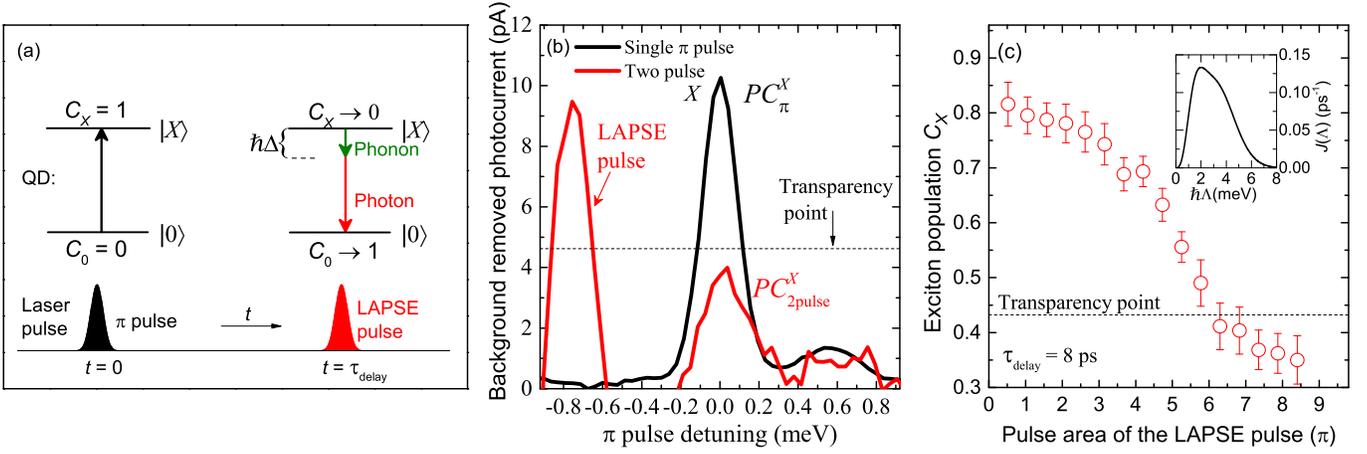}
	\caption{(a) Scheme of the two-pulse measurement. The QD is excited to the $|X\rangle$ state by a $\pi$ pulse at $t=0$ and then depopulated by a red-shifted laser pulse (LAPSE pulse) after $\tau_\text{delay}$.   (b) Black: PC spectrum measured only with a $\pi$ pulse  (FWHM = 0.2 meV, $\tau_\textbf{L}$ = 16.8 ps). Red: Two-pulse spectrum measured in the presence of a $-0.7$~meV detuned LAPSE pulse (FWHM = 0.42 meV, $\tau_\textbf{L}$ = 8 ps, $\Theta=5.25\pi$, $\tau_\text{delay}=7~\text{ps}$). $PC_\pi^{X}$ and  $PC_\text{2pulse}^{X}$ denote the amplitude of the exciton peak in the single-pulse and two-pulse spectra, respectively. The background of the PC spectra have been removed. (c)  Remaining exciton population $C_X$ after the LAPSE pulse versus the pulse area $\Theta$. LAPSE pulse: $\hbar\Delta=-0.7$~meV, FWHM = 0.2 meV,  $\tau_\textbf{L}$ = 16.8 ps, $\tau_\text{delay}=8$~ps. $C_X$ is deduced according to: $C_X = e^{-\tau_\text{delay}/\tau_\text{e}}-(1-PC^{X}_\text{2pulse}/PC^{X}_\pi)$, where $\tau_\text{e}$ is the electron tunnelling time [see details in Ref.~\onlinecite{SMnote}]. Inset: Calculated exciton-phonon coupling spectral density $J(\Lambda)$.}
	\label{fig6}
\end{figure*}

In the  LAPSE process, the QD is initially in the $|X_\text{R}\rangle$ state. By applying a red-shifted laser pulse ($\Delta<0$),  the phonon-assisted relaxation channel is activated and the system relaxes from the higher-energy more exciton-like state $|\alpha\rangle$ to the lower-energy more ground-state-like state $|\beta\rangle$ by emitting phonons with an energy of $\hbar \Lambda$ [see red arrow in Fig.~\ref{fig1}(d)]. If the phonon relaxation is fast enough, the system will reach thermal equilibrium between the two dressed states during the laser pulse. At low temperatures ($k_\text{B}T \ll \hbar\Lambda$) the lower dressed state $|\beta\rangle$ will be occupied predominantly and approaches $|0_\text{R}\rangle$ when the laser field is switched off \cite{Barth2016}. Inverting the sign of the laser detuning $\hbar\Delta$ leads to the opposite process: the creation of an exciton by absorbing a photon and emitting a phonon  \cite{Glassl2013, Hughes2013, Quilter2015, Ardelt2014, Bounouar2015} [see blue arrow in Fig.~\ref{fig1}(d)]. Here we note that LAPSE is fundamentally different from existing depopulation schemes, such as coherent control schemes \cite{Heberle1995, Zrenner2002} and adiabatic rapid passage protocols \cite{Wei2014, Mathew2014a, Luker2012} in which exciton-phonon coupling is usually a hindrance, whereas LAPSE is enabled by the exciton-phonon interaction.

To demonstrate the depopulation of a QD from above to below the transparency point, we first resonantly pump the QD to the exciton state at $t=0$ using a laser pulse with pulse area $\Theta=\pi$ [see Fig.~\ref{fig6}(a)].  $\Theta$ is defined as $(\mu_X/\hbar)\int_{-\infty}^{+\infty} E(t) \text{d}t$ and was determined from a Rabi oscillation measurement at $\hbar\Delta=0$~meV \cite{SMnote, Ramsay2010a, Ramsay2010c, Monniello2013}. $\mu_X$ is the optical dipole matrix element for the $|0\rangle\rightarrow|X\rangle$ transition. The black line in Fig.~\ref{fig6}(b) shows a PC spectrum measured as a function of the detuning of the $\pi$ pulse. The peak at zero detuning corresponds to the $|0\rangle\rightarrow|X\rangle$ transition. Its amplitude $PC^{X}_\pi$ corresponds to an exciton population of 1. Next, we apply a strong red-shifted pulse ($\hbar\Delta=-0.7$~meV, $\Theta=5.25\pi$) to depopulate the inverted QD. In the text below, we call this red-shifted laser pulse the ``LAPSE pulse". To maximize the efficiency of LAPSE, the width of the LAPSE pulse should be spectrally broad to cover as many phonon modes as possible, whereas the pulse duration $\tau_\textbf{L}$ needs to be long enough for the dressed states to complete the phonon-assisted relaxation. To fulfil both conditions, consistent with theory, we set the FWHM and $\tau_\textbf{L}$ of the LAPSE pulse to 0.42 meV and 8~ps. The red line in Fig.~\ref{fig6}(b) shows the PC spectrum measured in the presence of the LAPSE pulse. 
The peak at $\hbar\Delta=-0.7$~meV is due to the interference between the $\pi$ pulse and the LAPSE pulse. The reduction of the amplitude of the exciton peak $PC^{X}_\text{2pulse}$ relative to  $PC^{X}_\pi$ directly demonstrates the depopulation of the exciton state. In the PC measurement, the total population of the ground state and exciton state $C_\text{Total}$ drops to 0.88 at the arrival of the LAPSE pulse due to the electron tunnelling out from the QD during the delay time $\tau_\text{delay}=7$~ps; therefore the transparency point defined as $C_\text{Total}/2$ is shifted to 0.44 [see details in ref.~\onlinecite{SMnote}]. The fact that $PC^{X}_\text{2pulse}<0.44PC^{X}_\pi$ proves that the inverted QD is depopulated below the transparency point.

Fig.~\ref{fig6}(c) shows the remaining exciton population obtained after the LAPSE pulse versus the pulse area $\Theta$. In order to improve the spectral resolution and the signal to noise ratio, in the following measurements we reduce the FWHM of the LAPSE pulse to 0.2 meV. It can be seen that $C_X$ decreases with the increase of $\Theta$ as predicted by the simulation in Fig.~\ref{fig2}(b) and crosses the transparency point at $\Theta>6.25 \pi$. The minimum $C_X$ measured here is limited by the laser power available in our setup. The efficiency of the LAPSE process strongly depends on the pulse area because the laser detuning and power determine the effective Rabi splitting $\hbar\Lambda$  that needs to be in resonance with the phonon environment for an efficient phonon-assisted relaxation to occur \cite{Nazir2008}. Fig.~\ref{fig6}(c) inset shows the calculated exciton-phonon coupling spectral density $J$ as a function of the effective Rabi splitting: $J(\Lambda)=\sum_{\bf q} |\gamma_{\bf q}|^{2}\,\delta(\Lambda-\omega_{\bf q})$, where $\gamma_{\bf q}$  is the exciton-phonon coupling and $\bf q$ is the wave vector of bulk LA phonons [see details in Ref.~\onlinecite{SMnote}]. Here it becomes clear that the QD can be dynamically turned from a weakly vibronic system into a strongly vibronic one by laser driving when $\hbar\Lambda$ approaches 2~meV.

\begin{figure}[!h]
\includegraphics[width=\linewidth]{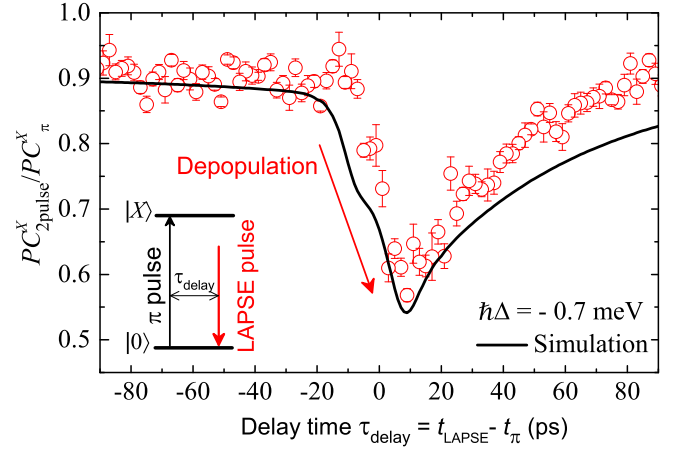}
\caption{Time dynamics of the LAPSE process. Red circles: $PC_\text{2pulse}^{X}/PC_\pi^{X}$ measured as a function of $\tau_\text{delay}$ under the same condition as Fig.~\ref{fig6}(c). $\tau_\text{delay}=t_\text{LAPSE}-t_{\pi}$, where $t_\text{LAPSE}$ and $t_{\pi}$ are the arrival times of the LAPSE pulse and the $\pi$ pulse. The pulse area of the LAPSE pulse is $5.25\pi$. Black line: Numerical simulation. Inset: Excitation scheme.}
\label{fig3}
\end{figure}

We next investigate the time dynamics of LAPSE by measuring $PC_\text{2pulse}^{X}/PC_\pi^{X}$ versus the delay time between the $\pi$ pulse and the LAPSE pulse [see red circles in Fig.~\ref{fig3}]. At negative delay time, the LAPSE pulse arrives before the exciton is created by the $\pi$ pulse and therefore cannot depopulate the exciton. The signal ($\sim0.9$) is not exactly 1 as the red-shifted LAPSE pulse can also create excitons with very small probability by absorbing phonons at $T>0$. When the $\pi$ pulse overlaps with the LAPSE pulse, $PC_\text{2pulse}^{X}$ decreases to a minimum from $\tau_\text{delay}=-10$ to $+10$~ps, indicating that the LAPSE process can be as fast as 20~ps. The signal then slowly recovers due to the electron tunnelling before the arrival of the LAPSE pulse. The electron tunneling time (55 ps) is measured using inversion recovery techniques \cite{Kolodka2007a}. The depopulation time is determined by the laser pulse width and in principle can be further reduced by using shorter laser pulses which however can diminish the efficiency of the LAPSE process when there is not enough time for the phonon-assisted relaxation to complete \cite{Glassl2013, Bounouar2015}. In ref.~\onlinecite{Glassl2013} it is found that a pulse duration of about $10$~ps is sufficient for the QD to reach thermal equilibrium between the two dressed states [see Fig.~\ref{fig1}(d)]. 

To further verify our understanding of the time dynamics of LAPSE, we have performed path-integral calculations based on a model of a laser-driven QD coupled to LA phonons [see details in ref.~\onlinecite{SMnote}]. The electron tunnelling occurring during the PC measurement is integrated as a Lindblad-type relaxation term into the path-integral approach without taking away the numerically complete treatment that includes all multi-phonon processes and non-Markovian effects. The calculation (black line) well reproduces all the features observed in the experiment, 
proving that the decrease of the PC signal is indeed caused by LAPSE.

By contrast to the fixed frequency of stimulated emission under resonant excitation, LAPSE can occur within a broad tuning range. To demonstrate this tunability, we measure the decrease of exciton population caused by a LAPSE pulse as a function of the laser detuning $\hbar\Delta$ [see Fig.~\ref{fig2}(a)]. In this measurement, a $\pi$ pulse creates a reference PC level [dashed line in Fig.~\ref{fig2}(a)] by resonantly pumping the QD. The reference level corresponds to an exciton population of 1. Then we apply a LAPSE pulse to depopulate the exciton state and measure the PC signal as a function of the laser detuning. To isolate the signal of the QD under study from other QDs in the sample, a reference spectrum is measured with only the LAPSE pulse and subtracted \cite{SMnote}. The colored lines in Fig.~\ref{fig2}(a) are the differential spectra $\Delta PC$ measured at different pulse areas. The dip at zero detuning corresponds to the resonant transition between $|X\rangle$ and $|0\rangle$. The reduction of the PC signal at negative detuning originates from the LAPSE process. The broad tuning range ($\sim$~4 meV) is clearly demonstrated by the negative sidebands. The good agreement between the experimental result and the path-integral calculations (black lines) supports our interpretation.

\begin{figure}
\centering
\includegraphics[width=\linewidth]{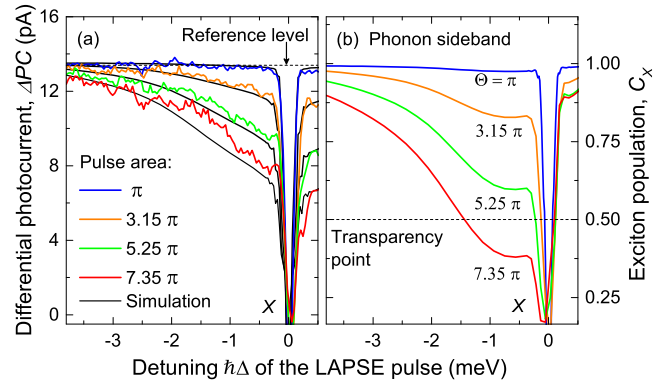}
\caption{ Tunability of the LAPSE process. (a) Diffirential photocurrent spectra measured by pumping the QD resonantly with a $\pi$ pulse and then depopulating the exciton by applying a LAPSE pulse at different detunings and pulse areas. To isolate the signal of the target QD from other QDs in the same sample, a reference spectrum where only the LAPSE pulse is applied is subtracted. $\tau_\text{delay} =$ 17~ps. Solid black lines: simulation. (b) Calculated phonon sideband without taking into account the electron tunneling and the subtraction of the reference spectra.}
\label{fig2}
\end{figure}

Although the shape and amplitude of the sidebands shown in Fig.~\ref{fig2}(a) are primarily determined by the spectral dependence of the exciton-phonon coupling shown in the inset in Fig.~\ref{fig6}(c), the differential spectra $\Delta PC$ are also slightly influenced by the exciton depopulation due to electron tunnelling and the shape of the subtracted reference spectra \cite{SMnote}. To obtain more insight into the spectral dependence of LAPSE, we calculate the remaining exciton population obtained after the LAPSE pulse without taking into account the electron tunnelling and subtraction of the reference spectra. This simulation directly shows how the probability of LAPSE depends on the laser detuning  [see Fig.~\ref{fig2}(b)]. Furthermore the simulation for $\Theta=7.35\pi$ clearly shows that the inverted QD is depopulated to below the transparency point at $\hbar\Delta\sim -0.7$~meV.

In conclusion, we have demonstrated the ultrafast depopulation of a QD from above to below the transparency point using red-shifted incoherent excitation via LA-phonon-assisted stimulated emission. The depopulation time ($\sim$20~ps) is determined by the laser pulse width, making it possible to reset a QD exciton much faster than the speed limitation imposed by the exciton lifetime. Due to the broadness of the phonon sideband, this scheme can occur in a tuning range of a few meV and may form the basis of tunable single QD lasers. Additionally, it can be potentially used for ultrafast optical switching \cite{Heberle1995, Bonadeo1998, Bose2012, Volz2012, Englund2012, Cancellieri2014}, semiconductor optical amplifiers \cite{Capua2014} and precisely controlling the emission time of a single photon source in cavities \cite{Heinze2015, Munoz2015, Portalupi2015, Muller2015}.

This work was funded by the EPSRC (UK) programme grant EP/J007544/1. A. M. Barth and V. M. Axt gratefully acknowledge the financial support from Deutsche Forschungsgemeinschaft via the Project No. AX 17/7-1. The authors thank  A. J. Ramsay for very helpful discussions and H. Y. Liu and M. Hopkinson for sample growth. 

\bibliography{References}

\end{document}


\title{Ultrafast Depopulation of a Quantum Dot by LA-phonon-assisted Stimulated Emission - Supplemental Material}

\author{F. Liu}
\author{ L. M. P. Martins}
\author{A. J. Brash}
\affiliation{Department of Physics and Astronomy, University of Sheffield, Sheffield, S3 7RH, United Kingdom}
\author{A. M. Barth}
\affiliation{Institut f\"{u}r Theoretische Physik III, Universit\"{a}t Bayreuth, 95440 Bayreuth, Germany}
\author{ J.~H.~Quilter}
\affiliation{Department of Physics and Astronomy, University of Sheffield, Sheffield, S3 7RH, United Kingdom}
\affiliation{Department of Physics, Royal Holloway, University of London, Egham, TW20 0EX, United Kingdom}
\author{V. M. Axt}
\affiliation{Institut f\"{u}r Theoretische Physik III, Universit\"{a}t Bayreuth, 95440 Bayreuth, Germany}
\author{M. S. Skolnick}
\affiliation{Department of Physics and Astronomy, University of Sheffield, Sheffield, S3 7RH, United Kingdom}
\author{A. M. Fox}
\affiliation{Department of Physics and Astronomy, University of Sheffield, Sheffield, S3 7RH, United Kingdom}

\maketitle

\section{Theoretical description of a driven QD}

To model the optically driven quantum dot (QD) coupled to longitudinal accoustic (LA) phonons we use the Hamiltonian \cite{Quilter2015}: 
\begin{align}
 \label{eq:Hamiltonian}
 H = H_{\rm{QD-light}} + H_{\rm{QD-phonon}},
\end{align}
where
\begin{align}
 H_{\rm{QD-light}} = \hbar\omega^0_{X}| X\rangle\langle X| + \quad\quad\quad\quad\quad\quad\quad\quad\quad\quad\quad\quad
 \\ \frac{\hbar\Omega(t)}{2}[| 0\rangle\langle X|e^{i \omega_L t} + |X\rangle \langle 0|e^{-i \omega_L t}], \quad\quad\quad
\end{align}
and
\begin{align}
 H_{\rm{QD-phonon}} \!=\! \sum_{\bf q} \hbar\omega_{\bf q}\,b^\dag_{\bf q} b_{\bf q}
\!+\! \sum_{\bf q} \hbar \big( \gamma_{\bf q} b_{\bf q} \!+\! \gamma^{\ast}_{\bf q} b^\dag_{\bf q}
                          \big) |X \rangle\langle X|.
\label{dot-ph}
\end{align}
$\hbar\omega^{0}_{X}$ denotes the transition energy between $|0\rangle$ and $|X\rangle$ states. $\Omega(t)$ denotes the Rabi frequency proportional to the electric field envelope of a circularly polarized laser pulse.
We refer to the difference between the laser frequency $\omega_{L}$ and 
the exciton energy by $\Delta = \omega_{L}-\omega_{X}$, where $\omega_{X}$ is the
frequency of the single exciton resonance deviating from $\omega^{0}_{X}$
by the polaron shift resulting from the dot-phonon coupling
in Eq.~(\ref{dot-ph}). $b^\dag_{\bf q}$ and $b_{\bf q}$ are the creation and annihilation operators for a LA phonon with wave vector $\bf{q}$ and energy $\hbar \omega_{\bf{q}}$. A linear dispersion relation $\omega_{\bf{q}} = c_{s} |\bf{q}|$ is assumed, where $c_{s}$ denotes the speed of sound.  The phonons are coupled to the QD via the deformation potential and the 
exciton-phonon coupling is expressed by 
$\gamma_{\bf{q}}=\frac{|\bf{q}|}{\sqrt{2V\rho \hbar \omega_{\bf{q}}}}
\left(D_{\rm{e}} \Psi^{\rm{e}}({\bf q}) - D_{\rm{h}} \Psi^{\rm{h}}({\bf
q})\right)$, where $\rho$ denotes the mass density of the crystal, $V$ the mode
volume, $D_{\rm{e/h}}$ the deformation potential constants, and
$\Psi^{\rm{e/h}}(\bf{q})$ the form factors of electron and hole, respectively.
We assume the system to be initially in a product state of a thermal phonon-distribution
at $4.2$~K and a pure ground-state of the electronic subsystem. For bulk LA phonons coupled via the deformation potential we obtain for a parabolic confinement \cite{Vagov2011}:
\begin{eqnarray}
J(\omega ) =\frac{\omega ^{3}}{4\pi ^{2}\rho \hbar v_{c}^{5}}
\left[
D_{\mathrm{e}}\,
e^{ \left( -\omega ^{2}a_{\mathrm{e}}^{2}/4v_{c}^{2} \right) }
-D_{\mathrm{h}}\,
e^{\left( -\omega ^{2}a_{\mathrm{h}}^{2}/4v_{c}^{2} \right) }
 \right]^{2}.
\label{eq_J}
\end{eqnarray}
We use the same material parameters given in ref.~\onlinecite{Quilter2015} for GaAs, which are: $\rho =
5370 \; \rm{kg}/\rm{m}^3$, $c_{s} = 5110 \; \rm{m}/\rm{s}$, $D_{\rm{e}} = 7.0 \;
\rm{eV}$, and $D_{\rm{h}} = -3.5 \; \rm{eV}$. The electron and
hole confinement lengths $a_{\mathrm{e/h}}$ are used as fitting parameters obtained from ref.~\onlinecite{Quilter2015}: $a_{\mathrm{e}}=4.5$~nm, $a_{\mathrm{h}}=1.8$~nm.

To also take into account the electron tunneling during the photocurrent (PC) measurements in the time-dependent path-integral simulation,
we add to the two QD levels $|0\rangle$ and $|X\rangle$ a third electronic level $|h\rangle$ accounting for the hole state remaining after an
electron tunnels out from the QD. The hole tunneling time is significantly longer than the measurement time and
therefore does not need to be included. This third level is hence only coupled to the exciton state $|X\rangle$ and the relaxation
from $|X\rangle$ to $|h\rangle$ is modelled as it is done in the Lindblad master equation approach \cite{Breuer2002}. The relaxation rate is given
by the measured electron tunneling time of 55 ps. Details of the incorporation of the Lindblad relaxation within
the path integral approach will be given elsewhere. When the simulation time is sufficiently long to complete
all relaxation processes the measured PC signal $PC^X_\text{2pulse}$ and $PC_\pi^X$ [see Fig. 3 in the main text] can be directly compared with the occupation of the $|h\rangle$
state, because it is a direct measure for the electrons tunneling out of the QD and hence must be proportional to the
induced PC.

\section{Determining the pulse area by Rabi oscillation measurement}

\begin{figure}[h]
\includegraphics[width=0.5\textwidth]{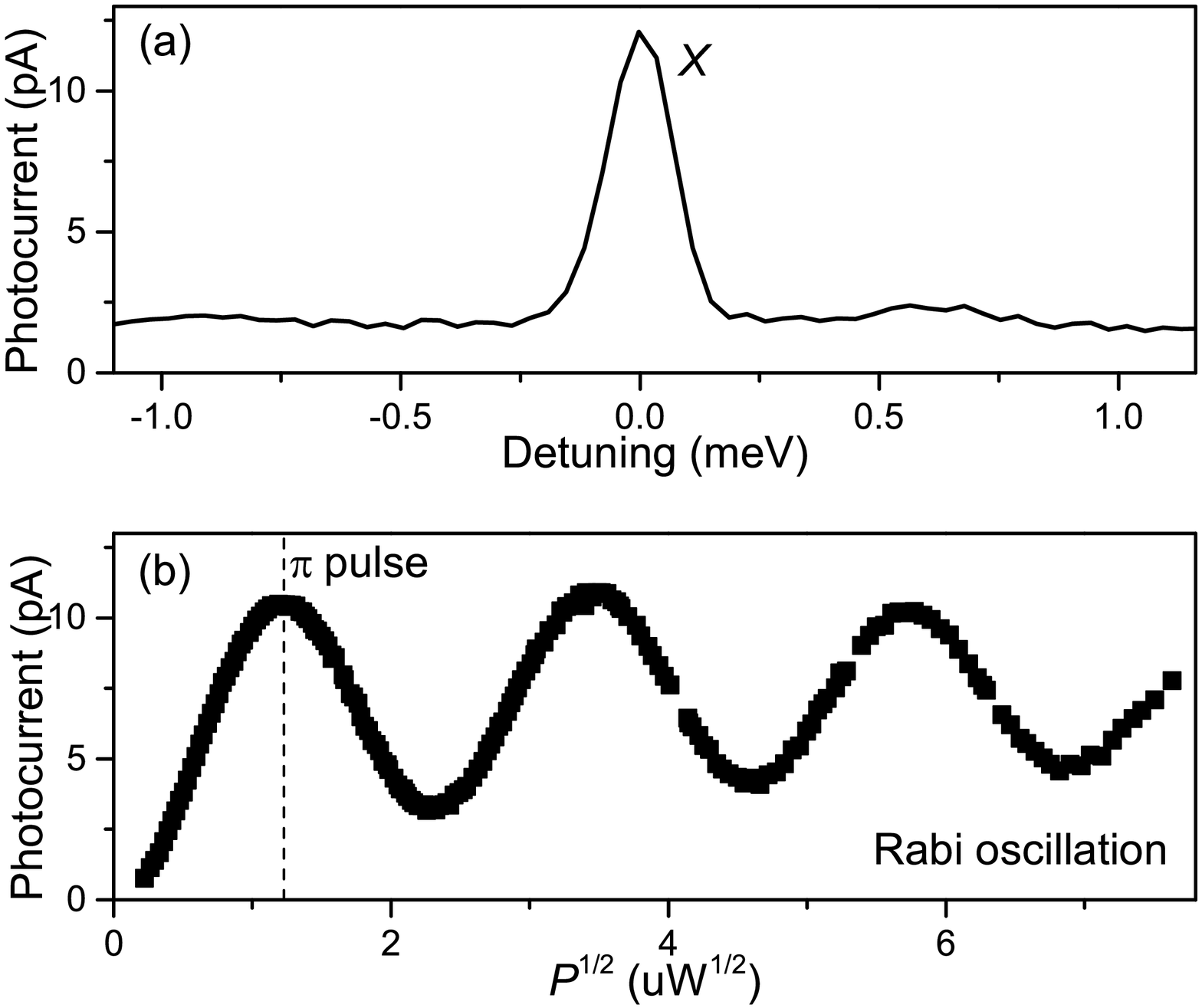}
\caption{(a) PC spectrum of a QD measured as a function of the $\pi$ pulse detuning. (b) Rabi oscillation measured as a function of the square root of the laser power at detuning = 0~meV.}
\label{fig0}
\end{figure}

The pulse area $\Theta$ of the laser pulse can be determined from the Rabi oscillation measurement. Fig.~\ref{fig0}(a) shows the PC spectra of a QD measured as a function of the laser detuning. The peak at 0 detuning corresponds to the neutral exciton transition. Then we measured the PC as a function of the laser power at detuning = 0~meV. The data in Fig.~\ref{fig0}(b) shows the Rabi oscillation of the exciton population after subtracting a PC background increasing linearly with the laser power. The laser power at the first maximum of the Rabi oscillation corresponds to a $\pi$ pulse. The peak amplitude of the PC spectrum of the QD measured with a $\pi$ pulse corresponds to an exciton population of 1.

\section{Effect of electron tunnelling on measured values of $C_X$}

In the two-pulse measurement shown in Fig.~2 in the main text, the remaining exciton population $C_X$ obtained after the red-shifted laser pulse (LAPSE pulse) is deduced in the following way. At $t=0$, the $\pi$ pulse creates an exciton population of 1. Then the exciton population decays exponentially since the electron tunnels out from the QD with a tunneling time $\tau_\text{e}$. The hole tunnelling and exciton radiative recombination are neglected because the hole tunnelling time $\tau_\text{h}$ (few ns) and the exciton radiative lifetime ($\sim600$~ps) \cite{Quilter2015} is significantly longer than $\tau_\text{e}$ (55~ps) and the delay time $\tau_\text{delay}$ (7-17~ps) between the $\pi$ pulse and the LAPSE pulses.  At $t=\tau_\text{delay}$ when the LAPSE pulse arrives, the exciton population is reduced from $e^{-\tau_\text{delay}/\tau_\text{e}}$ to $C_X$ by the LAPSE pulse via phonon-assisted stimulated emission. Since the amplitude of the measured PC signal is proportional to the time-integrated exciton  population created by the two pulses,  the change of the exciton population $\Delta C$ induced by the LAPSE pulse is proportional to the reduction of the PC signal $PC^X_\text{2pulse}$ measured in the presence of the LAPSE pulse relative to $PC^X_\pi$ measured with only a $\pi$ pulse [see Fig.~2(b) in the main text]:

\begin{eqnarray}
\Delta C=e^{-\tau_\text{delay}/\tau_\text{e}}-C_X=1-PC^{X}_\text{2pulse}/PC^{X}_\pi.
\end{eqnarray}
Hence we obtain:
\begin{eqnarray}
C_X = e^{-\tau_\text{delay}/\tau_\text{e}}-(1-PC^{X}_\text{2pulse}/PC^{X}_\pi).
\end{eqnarray}
The transparency point $C_\text{TP}(t)$ of the QD at $t=\tau_\text{delay}$ is defined as half of the total population of the ground state and exciton state, thus we have $C_\text{TP}(t=\tau_\text{delay})=e^{-\tau_\text{delay}/\tau_\text{e}}/2$.

\section{Differential photocurrent measurement}

\begin{figure}[h]
\includegraphics[width=\textwidth]{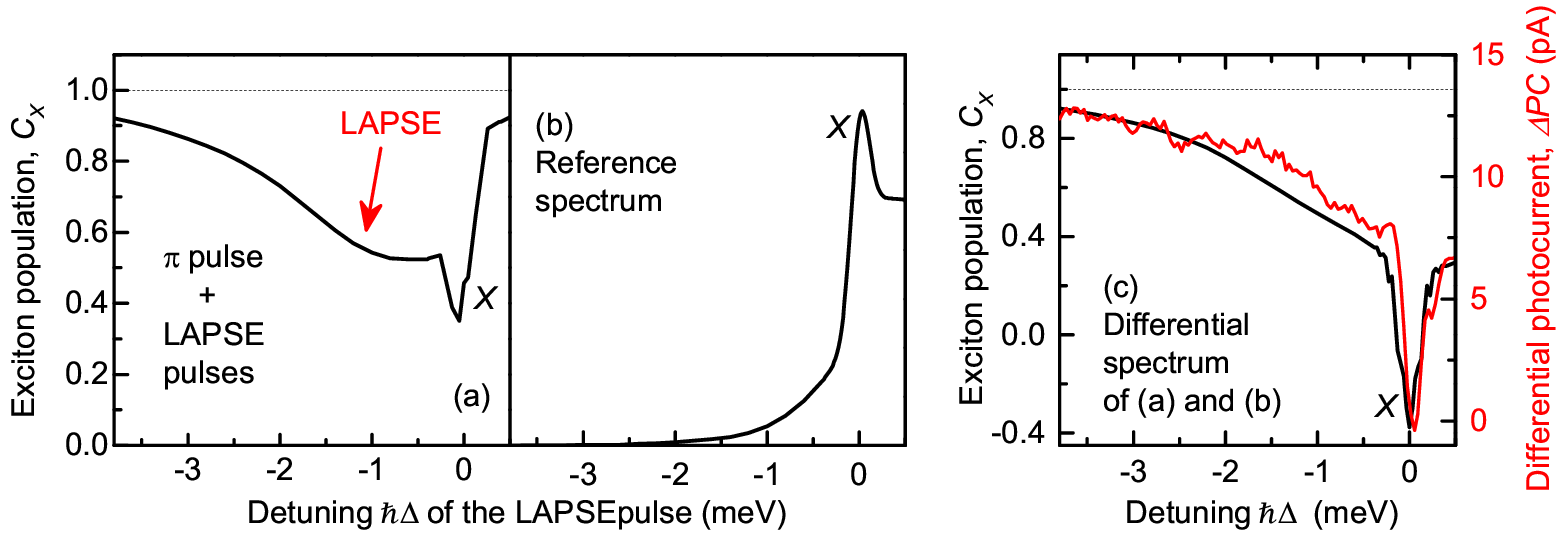}
\caption{ (a) - (c): Simulated single-QD spectra illustrating the steps of the differential PC measurement. (a) Exciton occupation as a function of the detuning of the LAPSE pulse which is applied after a resonant $\pi$ pulse and a delay time $\tau_\text{delay}$; (b) Exciton occupation obtained by scanning the detuning of only a LAPSE pulse; (c) Differential spectrum obtained by subtracting the spectrum in (b) from that in (a). Red line: measured differential PC spectra. $\Theta = 7.35~\pi$. $\tau_\text{delay} =$ 17~ps.}
\label{fig1}
\end{figure}

The differential PC spectra $\Delta PC$ shown in Fig. 4 in the main text were measured in two steps. In the first step, a two-pulse spectrum is obtained by pumping the QD to the $|X\rangle$ state using a circularly polarized resonant $\pi$ pulse and then measuring the PC as a function of the detuning of a strong laser pulse (LAPSE pulse) applied after a delay time $\tau_\text{delay}$. The exciton prepared by the $\pi$ pulse is expected to be deexcited via LAPSE when the depopulation pulse is negatively detuned ($\Delta<0$). This can be seen in Fig.~\ref{fig1}(a) which shows the simulated exciton occupation of a single QD obtained after the two pulses. However, in practice the phonon sideband in the negative detuning region is overlaid by the PC signal from nearby QDs in the same sample (not shown). In order to isolate the PC signal of the dot under study from that of the other QDs, in the second step a reference spectrum was measured by scanning the detuning of only a LAPSE pulse. The calculated exciton occupation is shown in Fig.~\ref{fig1}(b). Since the signals from the other QDs are present in both the two-pulse spectrum and the reference spectrum, they can be removed by subtracting the two spectra from each other. Fig.~\ref{fig1}(c) shows the measured  differential PC spectrum (red line) and the simulated spectrum (black line) obtained by subtracting the reference spectrum from the two-pulse spectrum. The calculated exciton population can be directly compared with the measured differential PC spectra by multiplying the exciton population with the amplitude of the exciton peak $PC_\pi^X$ measured with a $\pi$ pulse as shown in Fig. 4(a) in the main text. The sideband feature in the negative detuning region corresponds to the deexcitation of the exciton by the LAPSE process. 

\bibliography{References}